# Cubic anisotropy created by defects of "random local anisotropy" type, and phase diagram of the *O(n)* model

A.A. Berzin[1], A.I. Morosov[2]*, and A.S. Sigov[1]

[1] Moscow Technological University (MIREA), 78 Vernadskiy Ave., 119454 Moscow, Russian Federation
[2] Moscow Institute of Physics and Technology (State University), 9 Institutskiy per., 141700 Dolgoprudny, Moscow Region, Russian Federation

## Abstract

The expression for the cubic-type-anisotropy constant created by defects of "random local anisotropy" type is derived. It is shown that the Imry-Ma theorem stating that in space dimensions $d<4$ the introduction of an arbitrarily small concentration of defects of the "random local anisotropy" type in a system with continuous symmetry of the *n*-component vector order parameter (*O(n)* model) leads to the long-range order collapse and to occurrence of a disordered state, is not true if an anisotropic distribution of the defect-induced random easy axes directions in the order parameter space creates a global effective anisotropy of the "easy axis" type. For a weakly anisotropic distribution of the easy axes, in space dimensions $2 \leq d < 4$ there exists some critical defect concentration, when exceeded, the inhomogeneous Imry-Ma state can exist as an equilibrium one. At lower defect concentration the long-range order takes place in the system. For a strongly anisotropic distribution of the easy axes, the Imry-Ma state is suppressed completely and the long-range order state takes place at any defect concentration.

---
*E-mail: mor-alexandr@yandex.ru

## 1.  Introduction

Despite long-standing investigations of the systems with *n*-component vector parameter (*O(n)*-models), containing defects of the "random local anisotropy" type, we still have not a clear knowledge of their equilibrium properties. On the one hand, the validity of the Imry-Ma theorem [1] was carried over these systems and it was stated that introduction of an arbitrary small amount of defects of the "random local anisotropy" type leads to the long-range order collapse and to occurrence of a disordered state which hereafter we shall name the Imry-Ma state [2, 3].

On the other hand, a demand of the disorder isotropy was stated to be important for the Imry-Ma state realization [4].

It was shown in our preceding paper [5] that in the case when anisotropic distribution of the defect-induced random easy axes directions creates a global effective anisotropy of the "easy axis" type in the order parameter space it is necessary for the long-range order breakdown and the Imry-Ma state initiation that the constant of such a global anisotropy should not exceed its threshold value. Otherwise the long-range order does not die and the Imry-Ma state does not arise.

A set of anisotropic distributions of easy axes is not restricted to the case considered in Ref. [5] when a global anisotropy results from the straightforward summation of the energies of separate defects interaction with a homogeneous order parameter. An example of the distribution of easy axes differing from the isotropic one and, to a first approximation, not responsible for a global anisotropy is given by the distribution wherein local anisotropy axes are with equal probability directed parallel to *n* mutually perpendicular directions in the order parameter space which we choose as the axes of the Cartesian coordinate system.

The goal of the present paper is the study of the global anisotropy arising to the second and successive orders in the constant of local anisotropy created by a separate defect, and the development of the phase diagram of the system in "concentration of defects of "random local anisotropy" type – global anisotropy constant" variables.

2. **Energy of a system of classical spins**

The exchange-interaction energy of $n$-component localized unit (a vector length can be included to corresponding interaction or field constants) spins $\mathbf{s}_i$ comprising the simple cubic $d$-dimensional lattice, within the nearest neighbors approximation, has the form

$$W_{ex} = -\frac{1}{2} J \sum_{i,\delta} \mathbf{s}_i \mathbf{s}_{i+\delta}, \qquad (1)$$

where $J$ is the exchange interaction constant, the summation in $i$ is performed over the whole spin lattice, and the summation in $\delta$ is performed over the nearest neighbors.

The energy of interaction between the spins and "random local anisotropy" type defects is

$$W_{def} = -\frac{1}{2} K_0 \sum_l (\mathbf{s}_l \mathbf{n}_l)^2. \qquad (2)$$

Here $K_0 > 0$ is the random anisotropy constant, the summation is performed over defects randomly located in the lattice sites, and $\mathbf{n}_l$ is a unit vector prescribing the random easy axis direction.

Turning to the continuous distribution of the order parameter $\mathbf{s}(\mathbf{r})$, let introduce the inhomogeneous exchange energy in the form [6]

$$W_{ex} = \frac{D}{2} \int d^d \mathbf{r} \, \frac{\partial \mathbf{s}^\perp}{\partial x_i} \frac{\partial \mathbf{s}^\perp}{\partial x_i}, \qquad (3)$$

where $D = Jb^{2-d}$, $b$ is the interstitial distance, and $\mathbf{s}^\perp(\mathbf{r})$ is the order parameter component orthogonal to its mean direction $\mathbf{s}_0$.

The energy of interaction between spins and defects looks like

$$W_{def} = -\frac{1}{2} b^{-d} \int d^d \mathbf{r} \, K(\mathbf{r}) (\mathbf{s}(\mathbf{r}) \, \mathbf{n}(\mathbf{r}))^2, \qquad (4)$$

where

$$K(\mathbf{r}) = K_0 b^d \sum_l \delta(\mathbf{r} - \mathbf{r}_l). \tag{5}$$

### 3. Cubic type anisotropy

The anisotropy contribution to the volume energy density linear in the anisotropy constant $K_0$ is

$$w^{(1)}(\mathbf{s}_0) = -\frac{x\,K_0}{2b^d} \langle (\mathbf{s}_0 \mathbf{n}_l)^2 \rangle, \tag{6}$$

where $x$ is the dimensionless concentration of defects (the number of defects per a unit cell), and the brackets $\langle\ \rangle$ denote averaging over the whole set of defects. If the global anisotropy to the first order in $K_0$ is absent, that is subject to the condition $w^{(1)}(\mathbf{s}_0) = const$, one should take account of the order parameter inhomogeneity induced by random anisotropy and calculate the energy contribution quadratic (or higher power) of the constant $K_0$. Actually the expansion is performed in terms of a small parameter $K_0/J$. We neglect the longitudinal susceptibility of the system at low temperatures, much smaller than the temperature of magnetic ordering.

The presence of random anisotropy leads to a local deviation of the order parameter from its mean value and to the appearance of the component $\mathbf{s}^\perp(\mathbf{r})$ orthogonal to $\mathbf{s}_0$. The order parameter to the linear in $\mathbf{s}^\perp(\mathbf{r})$ approximation can be represented as

$$\mathbf{s}(\mathbf{r}) = \mathbf{s}_0 + \mathbf{s}^\perp(\mathbf{r}), \tag{7}$$

where $|\mathbf{s}_0| \gg |\mathbf{s}^\perp(\mathbf{r})|$. By substituting this expression to Eq. (4), we obtain linear in $\mathbf{s}^\perp(\mathbf{r})$ and quadratic in $K_0$ (as we shall see subsequently) summand to $W_{def}$

$$W^{(2)}_{def} = -b^{-d} \int d^d\mathbf{r}\, K(\mathbf{r})(\mathbf{s}_0 \mathbf{n}(\mathbf{r}))(\mathbf{s}^\perp(\mathbf{r})\mathbf{n}(\mathbf{r})). \tag{8}$$

The quantity $K(\mathbf{r})(\mathbf{n}(\mathbf{r})\,\mathbf{s}_0)\mathbf{n}(\mathbf{r})$ fulfils the role of an effective random field that acts on a spin. This field component $\mathbf{h}^\perp_{eff}(\mathbf{r})$ orthogonal to $\mathbf{s}_0$ is

$$\mathbf{h}_{eff}^{\perp}(\mathbf{r}) = K(\mathbf{r})\big(\mathbf{s}_0\mathbf{n}(\mathbf{r})\big)\big[\mathbf{n}(\mathbf{r}) - \mathbf{s}_0\big(\mathbf{s}_0\mathbf{n}(\mathbf{r})\big)\big]. \tag{9}$$

The Fourier component $\mathbf{s}^{\perp}(\mathbf{k})$ is related to the Fourier component of the effective random field $\mathbf{h}_{eff}^{\perp}(\mathbf{k})$

$$\mathbf{s}^{\perp}(\mathbf{k}) = \chi^{\perp}(\mathbf{k})\mathbf{h}_{eff}^{\perp}(\mathbf{k}), \tag{10}$$

where $\chi^{\perp}(\mathbf{k})$ is the Fourier component of the corresponding susceptibility of the spin system, and the quantity

$$\mathbf{h}_{eff}^{\perp}(\mathbf{k}) = \frac{1}{V}\int d^d\mathbf{r}\, \mathbf{h}_{eff}^{\perp}(\mathbf{r})\exp(-i\mathbf{k}\mathbf{r}) =$$

$$\frac{K_0}{N}\sum_l (\mathbf{s}_0\mathbf{n}_l)[\mathbf{n}_l - \mathbf{s}_0(\mathbf{s}_0\mathbf{n}_l)]\exp(-i\mathbf{k}\mathbf{r}_l), \tag{11}$$

where $V$ is the system volume, and $N$ is the number of elementary cells. Then one has

$$\mathbf{s}^{\perp}(\mathbf{r}) = \frac{K_0}{N}\sum_{\mathbf{k}}\chi^{\perp}(\mathbf{k})\sum_l(\mathbf{s}_0\mathbf{n}_l)[\mathbf{n}_l - \mathbf{s}_0(\mathbf{s}_0\mathbf{n}_l)]\exp[i\mathbf{k}(\mathbf{r}-\mathbf{r}_l)], \tag{12}$$

the summation is performed over all $\mathbf{k}$ from the Brillouin zone. Substitution of this expression into Eq. (8) gives

$$W_{def}^{(2)} = -\frac{K_0^2}{N}\sum_{\mathbf{k}}\chi^{\perp}(\mathbf{k})\sum_{l,m}(\mathbf{s}_0\mathbf{n}_l)[\mathbf{n}_l - \mathbf{s}_0(\mathbf{s}_0\mathbf{n}_l)](\mathbf{s}_0\mathbf{n}_m)[\mathbf{n}_m - \mathbf{s}_0(\mathbf{s}_0\mathbf{n}_m)]$$

$$\times \exp[i\mathbf{k}(\mathbf{r}_m - \mathbf{r}_l)]. \tag{13}$$

By virtue of a random distribution of defects in the coordinate space and a random choice of defect-induced local anisotropy axes, a nonzero contribution to $W_{def}$ results from the summands with $l=m$. Therefore Eq. (13) yields

$$W_{def}^{(2)} = -\frac{K_0^2}{N}\sum_{\mathbf{k}}\chi^{\perp}(\mathbf{k})\sum_l(\mathbf{s}_0\mathbf{n}_l)^2[1-(\mathbf{s}_0\mathbf{n}_l)^2] =$$

$$- xK_0^2\sum_{\mathbf{k}}\chi^{\perp}(\mathbf{k})[\langle(\mathbf{s}_0\mathbf{n}_l)^2\rangle - \langle(\mathbf{s}_0\mathbf{n}_l)^4\rangle]. \tag{14}$$

If $w^{(1)}(\mathbf{s}_0) = const$, then the second summand in the right-hand side of Eq. (14) describes the global anisotropy of the system.

For space dimensions 2<d<4, one can use the susceptibility of the pure system for the quantity $\chi^\perp(\mathbf{k})$

$$\chi^\perp(\mathbf{k}) = (Jb^2k^2)^{-1}. \tag{15}$$

Going from summation over $\mathbf{k}$ to integration over the Brillouin zone and introducing the notation

$$\tilde{\chi}^\perp = \int \frac{d^d\mathbf{k}}{(2\pi)^d} \chi^\perp(\mathbf{k}), \tag{16}$$

we get the quadratic in $K_0$ contribution to volume density of the interaction energy

$$w_{def}^{(2)} = -xK_0^2\tilde{\chi}^\perp[\langle(\mathbf{s}_0\mathbf{n}_l)^2\rangle - \langle(\mathbf{s}_0\mathbf{n}_l)^4\rangle]. \tag{17}$$

For space dimensions 2<d<4, the quantity $\tilde{\chi}^\perp$ has no peculiarities at $\mathbf{k} = 0$.

The quadratic in $K_0$ contribution of the inhomogeneous exchange energy (3) to the volume energy density can be found in the similar way, by substituting the expression for $\mathbf{s}^\perp(\mathbf{r})$ into Eq. (3). Such a contribution appears to be of the opposite sign and one half in magnitude as to $w_{def}^{(2)}$ given by Eq. (17). Therefore the resulting volume density of the anisotropy energy takes the form

$$w^{(2)} = -\frac{x\tilde{\chi}^\perp K_0^2}{2}[\langle(\mathbf{s}_0\mathbf{n}_l)^2\rangle - \langle(\mathbf{s}_0\mathbf{n}_l)^4\rangle]. \tag{18}$$

The summands containing higher powers of $K_0$ can be derived along similar lines.

The following value is taken as a global anisotropy constant $K_{eff}$

$$K_{eff} = 2b^d\left(w_{max}^{(1)} + w_{max}^{(2)} - w_{min}^{(1)} - w_{min}^{(2)}\right), \tag{19}$$

where $w_{max}^{(i)}$ and $w_{min}^{(i)}$ (i=1,2) are maximum and minimum values of $w^{(i)}$ as a function of vector $\mathbf{s}_0$ direction.

## 4. Two-dimensional model

A specific feature of two-dimensional models is the absence of the long-range order in a pure system at finite temperature, and so one has to anticipate the existence of the long-range order induced by random local anisotropy axes and solve a self-consistent problem [7].

Since under the influence of random field the order parameter deviates from the easy direction to the hard one, the expression for $\chi^\perp(\mathbf{k})$ takes the form

$$\chi^\perp(\mathbf{k}) = \left(Jb^2k^2 + K_{eff}\right)^{-1}. \qquad (20)$$

It can be easily seen that $K_{eff}$ cuts the divergence of $\tilde{\chi}^\perp$ at small $\mathbf{k}$ values which bring the main contribution to $\tilde{\chi}^\perp$ for $d=2$. As the result we obtain

$$\tilde{\chi}^\perp = \frac{1}{4\pi b^2 J} \ln \frac{4\pi J}{K_{eff}}. \qquad (21)$$

The value of $K_{eff}$ can be found by solving the self-consistency equation (19) after substituting the value of $\tilde{\chi}^\perp$. Since, to the first approximation, one has $K_{eff} \propto x$, the next approximation yields $w^{(2)} \propto -x\ln x$. Thus, despite the fact that the summand $w^{(2)}$ contains additional small parameter $K_0/J$ compared to $w^{(1)}$, its value can exceed $w^{(1)} \propto x$ in the region of low concentrations. An analysis of cubic and higher order terms in $K_0/J$ parameter demonstrates that at low concentrations their concentration dependence is the same as that of the term $w^{(2)}$. Consequently, when considering the global anisotropy concentration dependence, we can restrict ourselves to the summands linear and quadratic in $K_0$.

## 5. An example of cubic anisotropy

By way of illustration let us consider the case when random easy axes of defects are with equal probability directed collinearly to the Cartesian coordinate system axes in the *n*-dimensional space of the order parameter. In such a case $\langle n_{li} n_{lj} \rangle = \frac{1}{n}\delta_{ij}$, where $n_{lj}$ is *j*-th component of vector $\mathbf{n}_l$; *i,j*=1, 2, …, *n*; $\delta_{ij}$ is the Kroneker symbol, and $w^{(1)}(\mathbf{s}_0) = const$. Similarly, one finds $\langle n_{lj}^4 \rangle = \frac{1}{n}$, mean products of different components of вектора $\mathbf{n}_l$ equal zero, and $\langle (\mathbf{s}_0 \mathbf{n}_l)^4 \rangle = \frac{1}{n}\sum_{j=1}^{n} s_{0j}^4$. Maximum value of the sum $\sum_{j=1}^{n} s_{0j}^4$ equals 1 if vector $\mathbf{s}_0$ is parallel to an axis of the Cartesian coordinate system in the order parameter space, and its minimum value is $1/n$ when this vector is directed along one of the main diagonals of the given system.

Since the summand containing the term $\langle (\mathbf{s}_0 \mathbf{n}_l)^4 \rangle$ in the volume energy density is plus in sign, in the equilibrium state vector $\mathbf{s}_0$ is directed along one of the main diagonals of the Cartesian coordinate system in the order parameter space. A number of the order parameter components ($n=2$ in the X-Y model or $n=3$ in the Heisenberg model) does not play a significant part, because in contrast to the anisotropy linear in $K_0$, the anisotropy of the second and higher powers in $K_0$ leads to the occurrence of easy axes but not easy planes.

The global anisotropy constant has the form

$$K_{eff} = x\tilde{\chi}^\perp b^d K_0^2 \frac{n-1}{n^2}. \tag{22}$$

For the space dimension $2<d<4$ one has $\tilde{\chi}^\perp b^d J = const$ (for $d=3$ this value is approximately 0.2) and $K_{eff}$ can be evaluated as

$$K_{eff} \sim 0.1 x \frac{K_0^2}{J}. \tag{23}$$

For $d=2$, by using Eqs. (21), (22) and solving the self-consistency equation (19), we obtain to the logarithmic approximation

$$K_{eff} = x \frac{K_0^2}{4\pi J} \frac{n-1}{n^2} \ln \frac{(4\pi J n)^2}{x(n-1)K_0^2}. \tag{24}$$

## 6. Phase diagram of the system

An anisotropic distribution of random easy axes of defects induces anisotropy of both "easy axis" and "easy plane" types. The Imry-Ma inhomogeneous state is suppressed only by the "easy axis" type anisotropy [5, 8]. That is why we detail the "easy plane" type anisotropy case.

Gaining an answering the question if there arises in the system the long-range order with vector $s_0$ in the easy plane or the Imry-Ma inhomogeneous state, one should project all random vectors $n_l$ onto the given m-dimensional ($n>m\geq 2$) hyperplane in the order-parameter space and treat the problem at this hyperplane. When the "easy plane" anisotropy arises, the operation should be repeated. As a result we arrive at three possible cases:

(1) projections of random vectors $n_l$ on the easy plane equal zero. The system behavior therewith is analogous to that of the pure system with the number of the order parameter components corresponding to the hyperplane dimensionality. In any event the Imry-Ma inhomogeneous state does not occur;

(2) the "easy axis" anisotropy takes place in the easy plane itself. Then the problem reduces to that with the given anisotropy, but the number of the order parameter components equals $m$;

(3) the distribution of vectors $n_l$ projections on the easy plane is perfectly isotropic. In this case the Imry-Ma theorem is true.

In order to understand if the Imry-Ma inhomogeneous state is realized in a given system with the "easy axis" effective anisotropy type, it is necessary that the effective anisotropy constant be confronted with its critical value, wherein the state in question is suppressed [8, 5]. Indeed, to follow the space fluctuations of the easy axis direction, the order parameter has to deviate from the global easy axis direction. This leads to an increase in the volume density of the anisotropy energy by the value of the order of $K_{eff}/b^d$. When such a growth is not compensated by the gain in energy due to the order parameter alignment with the fluctuations of the easy axis direction, the Imry-Ma inhomogeneous state becomes energetically unfavorable, and the system goes back to the state with the long-range order.

The requisite critical value was found in our earlier papers [8, 5]:

$$K_{cr} \sim K_0 x^{\frac{2}{4-d}} \left(\frac{K_0}{J}\right)^{\frac{d}{4-d}}. \tag{25}$$

For space dimensionality 2<d<4, the global anisotropy induced by random defects is proportional to defect concentration $x$, while the quantity $K_{cr}$ contains a higher power of defect concentration. In particular, for $d=3$, one has $K_{cr} \propto x^2$. It follows from this that in the limit $x \to 0$ the effective anisotropy arising in any power to $K_0$ is bound to exceed its critical value.

If $d=2$, one has $K_{eff} \propto -x \ln x$, that is, in the region of small concentration the value of $K_{eff}$ also exceeds its critical value $K_{cr} \propto x$.

Thus the Imry-Ma theorem ceases to be valid at any arbitrarily small effective anisotropy of the "easy axis" type induced by random local anisotropy axes of defects. In the case of strongly anisotropic distributions of random easy axes, the Imry-Ma state does not occur in the whole range of defect concentrations $x < 1$.

For slightly anisotropic distributions of random local easy axes, the condition $K_{eff} < K_{cr}$ imposes a lower limit to the concentration of defects at which the Imry-Ma inhomogeneous state takes place [5].

The phase diagram characteristic of the system is displayed in Fig. 1.

## 7. Conclusions

An anisotropic distribution of defect-induced random local easy axes directions initiates the effective anisotropy of either "easy axis" or "easy plane" type in the order parameter space. The Imry-Ma theorem stating that at space dimensions $d<4$ the introduction of an ***arbitrarily small concentration*** (italicized by the present authors) of defects of the "random local anisotropy" type in a system with continuous symmetry of the $n$-component vector order parameter ($O(n)$ model) leads to the long-range order collapse and to the occurrence of a disordered state, breaks down at the advent of the "easy axis" anisotropy induced by the defects designed initially for breaking down the long-range order.

In the case of slightly anisotropic distribution of local easy axes, there exists a critical concentration of defects, above which the Imry-Ma inhomogeneous state can exist as an equilibrium one.

In the case of strongly anisotropic distribution of local easy axes, the Imry-Ma inhomogeneous state is completely suppressed, and the state with the long-range ordering is realized at any defect concentration.

## Figure caption

1. Phase diagram of the system in variables "inverse constant of the "easy axis" type effective anisotropy $K_{eff}^{-1}$ - concentration of defects $x$": LRO denotes the phase with the long-range order; I-M denotes the Imry-Ma inhomogeneous phase.

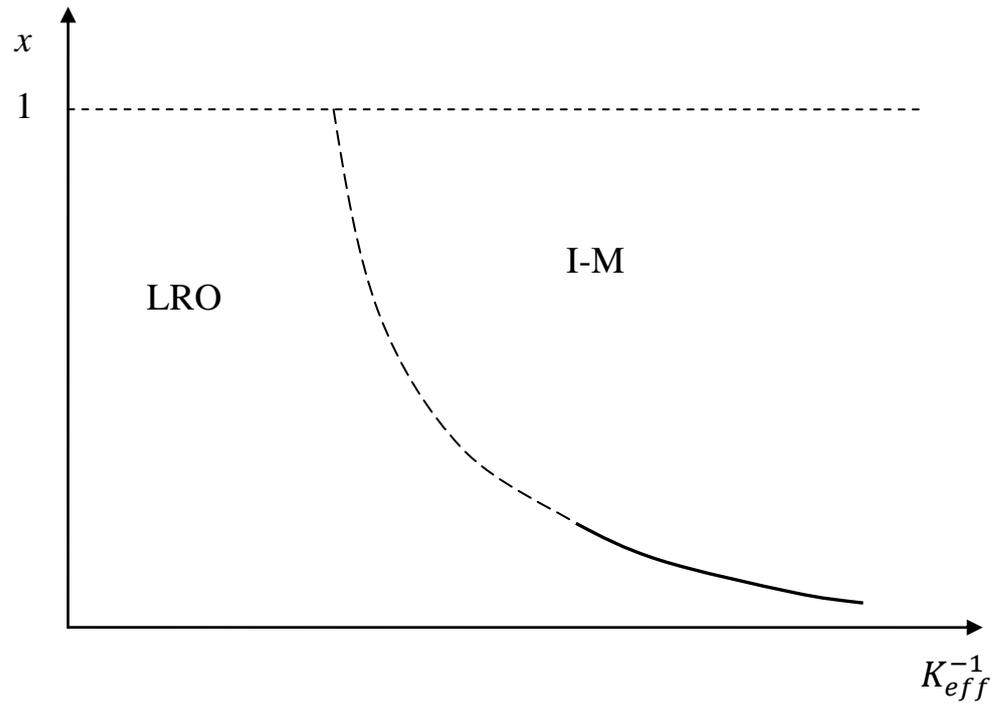